\documentclass[usegraphicx]{mn2e}

\usepackage{color}

\begin{document}
\def\mpc{h^{-1} {\rm{Mpc}}}
\def\kpc{h^{-1} {\rm{kpc}}}
\def\up{h^{-3} {\rm{Mpc^3}}}
\def\uk{h {\rm{Mpc^{-1}}}}
\def\lsim{\mathrel{\hbox{\rlap{\hbox{\lower4pt\hbox{$\sim$}}}\hbox{$<$}}}}
\def\gsim{\mathrel{\hbox{\rlap{\hbox{\lower4pt\hbox{$\sim$}}}\hbox{$>$}}}}
\def\kms {\rm{km~s^{-1}}}
\def\masa{h^{-1}{{\cal M}_{\odot}}}
\def\apj {ApJ}
\def\aj {AJ}
\def\aa {A \& A}
\def\mnras {MNRAS}
\definecolor{darkred}{rgb}{0.5,0.0,0.0}
\definecolor{darkgreen}{rgb}{0.0,0.3,0.0}
\newcommand{\new}[1]{{\textcolor{blue}{ #1}}}
\newcommand{\manolis}[1]{{\textcolor{green}{ #1}}}
\newcommand{\cinthia}[1]{{\it\textcolor{red}{ #1}}}
\newcommand{\nota}[1]{{\bf\textcolor{darkgreen}{DELETE ?: #1}}}
\newcommand{\mincir}{\raise
-2.truept\hbox{\rlap{\hbox{$\sim$}}\raise5.truept\hbox{$<$}\ }}
\newcommand{\magcir}{\raise
-2.truept\hbox{\rlap{\hbox{$\sim$}}\raise5.truept\hbox{$>$}\ }}
\newcommand{\minmag}{\raise
-2.truept\hbox{\rlap{\hbox{$<$}}\raise6.truept\hbox{$<$}\ }}

\title{The Relation Between Halo Shape, Velocity Dispersion and Formation Time}
\author[]
{
  \parbox[t]{\textwidth}
  {
    C. Ragone-Figueroa${^{1,2}}$, M. Plionis${^{3,4}}$, M. Merch\'an${^{1,2}}$,
    S. Gottl\"ober$^5$ and G. Yepes$^6$
  }
  \vspace*{6pt}\\ 
  \parbox[t]{15 cm}
  {
    $1$ Instituto de Investigaciones en Astronom\'{\i}a Te\'orica y 
    Experimental, 
    IATE, Observatorio Astron\'omico, Laprida 854, 5000, 
    C\'ordoba, Argentina.\\ 
    $2$ Consejo de Investigaciones Cient\'{\i}ficas y T\'ecnicas de la
    Rep\'ublica Argentina.\\
    $3$ Institute of Astronomy \& Astrophysics, National Observatory of Athens,
    Palaia Penteli 152 36, Athens, Greece.\\
    $4$ Instituto Nacional de Astrof\'{\i}sica Optica y Electr\'onica, AP 51
    y 216, 72000, Puebla, M\'exico.\\
    $5$ Astrophysical Institute Potsdam, An der Sternwarte 16, 
    14482 Potsdam, Germany\\
    $6$ Grupo de Astrof\'\i sica, Departamento de F\'isica Te\'orica, M\'odulo C-XI,
     Universidad Aut\'onoma de Madrid, Cantoblanco E-280049, Spain\\
  }
}
\date{\today}

\maketitle

\title{Morphology \& Dynamics of Group Size Haloes}

\begin{abstract}
We use dark matter haloes identified in the {\sc MareNostrum Universe} 
and galaxy groups identified in the Sloan Data Release 7 galaxy catalogue,
to study the relation between halo shape and halo dynamics, 
parametrizing out the mass of the systems.
A strong shape-dynamics, independent of mass, correlation is present in the simulation
data, which we find it to be due to different halo formation times. 
Early formation time haloes are, at the present epoch, more 
spherical and have higher velocity dispersions than late forming-time
haloes. The halo shape-dynamics correlation, albeit weaker, survives the projection in
2D (ie., among projected shape and 1-D velocity dispersion).
A similar shape-dynamics correlation, independent of mass, 
is also found in the SDSS DR7 groups of galaxies and in order to investigate
its cause we have tested and used, as a proxy of
the group formation time, a concentration parameter.
We have found, as in the case of the simulated haloes, 
that less concentrated groups, corresponding to late formation times, 
have lower velocity dispersions and higher elongations than groups with higher values of concentration,
corresponding to early formation times.

\end{abstract}


\section{INTRODUCTION}

\label{intro}

According to the hierarchical model of structure formation, groups and
clusters of galaxies, embedded in dark matter haloes, 
emerge from Gaussian primordial density fluctuations and grow by
accreting smaller structures, formed earlier, 
along anisotropic directions. Such structures constitute therefore an
important step in the 
hierarchy of cosmic structure formation and are extremely
important in deciphering the processes of galaxy and structure
formation. 
It should thus be expected that recently formed structures are highly
elongated, reflecting exactly the anisotropic distribution of the
large-scale structure from which they have accreted their mass, and
dynamically young.
Therefore the determination of the dynamical state of groups and
clusters of galaxies and
its evolution, which could be affected by a multitude of factors - like
environment, formation time, etc -  is a fundamental step
in investigating the hierarchical galaxy formation theories.

Both analytical calculations and N-body simulations have 
consistently shown that the virialization process will tend to
sphericalize the initial anisotropic distribution of matter, and
therefore the shape of haloes is an indication of their evolutionary
stage. Furthermore, the halo shape, size, and velocity dispersion
(among others)
are important factors in determining halo member orbits and
interaction rates, which are instrumental in understanding galaxy
formation and evolution processes (eg., Jing \& Suto 2002; Kasun \&
Evrard 2005; 
Avila-Reese et al. 2005; Bailin \& Steinmetz 2005; Allgood
et al. 2006; Bett et al. 2007; Ragone-Figueroa \& Plionis 2007; Macci\'o, Dutton, van
den Bosch 2008 and references therein). 
It has been found that dark-matter haloes are triaxial but with a 
strongly preferred prolatness (eg., Plionis, Basilakos \&
Ragone-Figueroa 2006; Gottl\"ober \& Yepes 2007 and references therein) 
and that there is also a
correlation between the orientation of their major axes and the 
surrounding structures. Such alignment effects,
among relatively massive haloes, have been found to be 
particularly strong (eg., Faltenbacher et al. 2002; Kasun \& Evrard 2005; 
Hopkins, Bachall \& Bode 2005; Ragone-Figueroa \& Plionis 2007; 
Faltenbacher et al. 2008 and references therein).
There are indications that the mentioned 
correlation might arise, or strengthen, from re-arrangements of the orientation of the halo 
axes in the direction from which the last major merger event occurred 
\cite{van}. 
Hence, the shape of dark matter haloes might well be correlated with the 
large scale
structure, providing clues to the fashion in which mass is aggregated
(eg., Basilakos et al. 2006). 
The imprints of this accretion can be observed in the substructure features
present in a halo, which have been found to correlate with the halo shape 
(Ragone-Figueroa \& Plionis 2007). Haloes with higher levels of substructure,
or equivalently dynamically young haloes, are preferentially more
elongated. The subsequent evolution should induce a change in
the shape of haloes, leading to more spherical configurations,
increasing their velocity dispersions as well.
 
According to this last statement, Faltenbacher \& Mathews (2007) have 
found, in agreement with the Jeans equation, that the velocity dispersion 
of sub-haloes increases with the host halo concentration. 
Given that halo concentration and formation time are also correlated 
(Bullock et al. 2001; Wechsler et al. 2002), 
this result implies that older haloes should have higher 
sub-halo velocity dispersions. 
Once a halo is virialized, the dynamical mass can be computed from their
velocity dispersion and virial radius (eg., Binney \& Tremaine 1987). 
Nevertheless, if virialization has not yet been achieved, 
the above mentioned effect and its possible dependence to the local halo environment
should be taken into account.
In addition, there is also evidence for the 
sphericalization of DM haloes with time (e.g. Allgood et al. 2006).
This behavior can be well studied using haloes in numerical 
simulations, where their 3D shape and velocity dispersion can be calculated 
accurately. 

These issues motivated us to study the relation between the DM halo 
velocity dispersion and shape, in narrow halo mass ranges, in order to exclude
the possible halo-mass dependent effects.
We also investigate how this correlation is transformed when 
computing projected shapes
and velocity dispersion along the line-of-sight, in order to be able
to compare
with observational data, like the SDSS galaxy groups.

The plan of the paper is the following. In section \ref{datos} we
describe the {\sc MareNostrum} numerical simulation and the SDSS 
observational data together with
the identification procedure of the corresponding systems and their properties.
In section \ref{hsdr} we study the halo shape-dynamics relation in the
simulation. In section \ref{proxies} we seek for a formation-age proxy
which can be computed in a realistic group catalog. We continue
in section \ref{2d} by presenting the shape-dynamics correlation results
for the observed SDSS groups of galaxies.
Finally, section \ref{conc} contains our conclusions.

\section{THE HALO \& GROUP DATA}
\label{datos}
\subsection{The {\sc MareNostrum} Simulation}

This non-radiative SPH simulation, dubbed {\sc The MareNostrum
Universe} (see Gottl\"ober \& Yepes 2007) and covering a volume of $(500 \; h^{-1} {\rm
Mpc})^3$, was performed in 2005 with
the parallel TREEPM+SPH {\sc Gadget2} code (Springel 2005). 
The resolution of the simulation is such that the gas and the DM
components are resolved by $2 \times
1024^3$ particles. The initial conditions at redshift $z=40$ were
calculated assuming a spatially flat concordance cosmological model
with the following parameters: the total matter density
$\Omega_m=0.3$, the baryon density $\Omega_b=0.045$, the cosmological
constant $\Omega_\Lambda=0.7$, the Hubble parameter $h=0.7$, the slope
of the initial power spectrum $n=1$ and the normalization
$\sigma_8=0.9$.  This resulted in a mass of $8.3 \times 10^9\masa$ for
the DM particles and $1.5 \times 10^9\masa$ for the gas particles. The
simulation followed the nonlinear evolution of structures in gas and
dark matter from $z=40$ to the present epoch. Dissipative or radiative
processes or star formation were not included.  The spatial force
resolution was set to an equivalent Plummer gravitational softening of
$15 \;h^{-1}$ comoving kpc.  The SPH smoothing length was set to the
distance to the 40$^{th}$ nearest neighbor of each SPH particle.

After the release of the 3-year WMAP data the original simulation has been
repeated with lower resolution assuming the predicted low WMAP3
normalization ($\sigma_8=0.75$) as well as with a higher normalization
of $\sigma_8 = 0.8$, which is better in agreement with the 5-year WMAP
data. Yepes at al. (2007) argue that the low normalization
cosmological model inferred from the 3 year WMAP data results is
barely compatible with the present epoch X-ray cluster abundances. All
these simulations have been performed within the MareNostrum Cosmology
project at the Barcelona Supercomputer Center.

\subsubsection{Halo identification, shape \& formation time} 

In order to find all structures and substructures within the
distribution of 2 billion particles and to determine their properties
we have used a hierarchical friends-of-friends algorithm (Klypin et
al. 1999, Gottl\"ober et al. 2006b). At all redshifts we have used a
basic linking length of 0.17 of the mean inter-particle separation to
extract the FOF clusters, which corresponds to an overdensity of 
$\sim$330 of the mean density. Shorter linking lengths have been used to
study substructures. The FOF analysis has been performed independently
over DM and gas particles. Using a linking length of 0.17 at redshift
$z=0$ we have identified more than 2 million objects with more than 20
DM particles which closely follow a Sheth-Tormen mass function
(Gottl\"ober et al. 2006a).

Using the FOF method one extracts rather complex objects from the
simulation which are characterized by an iso-density surface given by
the linking length. 
In first approximation these objects can be
characterized by triaxial ellipsoids. The shape and orientation of
the ellipsoids can be directly calculated as eigenvectors of the
inertia tensor of the given object. Then the shape is characterized by
the ratios between the lengths of the three main axes
($a>b>c$). Gottl\"ober \& Yepes 2007 have studied the shape of the
galaxy clusters in the {\sc The MareNostrum Universe} and found that
both the gas and dark matter components tend to be prolate although
the gas is much more spherically distributed. Here we study the shape
of the more massive dark matter component of the FOF objects which
reflects the shape of the potential in which the galaxies move.

Finally, in our present work we will extensively use the notion of 
halo formation time ($z_{\rm form}$),
which is usually defined as the redshift at which the halo accretes half of
its final ($z=0$) mass. 
By selecting haloes in the lower and upper quartiles of the formation time 
distribution, 
we define what we will call the late formation time (LFT) and early formation 
time (EFT) halo populations, respectively.
We also divide the haloes in four mass ranges, namely
$1.0-1.3$, $2.0-2.5$, $3.2- 4.0$ and $6.3-7.9 \times10^{13} \masa$,
since we wish to cancel the dependence of any halo-based parameter on
its mass.

\begin{figure}
\centering
\resizebox{\hsize}{!}{\includegraphics{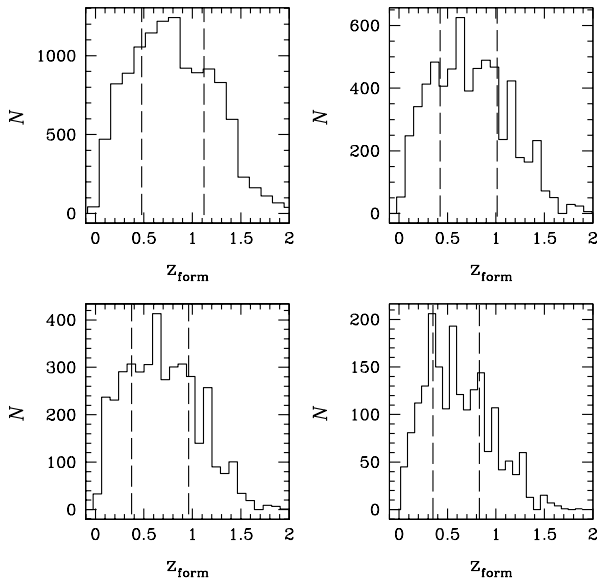}}
\caption{ 
Different panels correspond to the formation time distribution of
haloes in a different range of mass, namely
$1.0-1.3$, $2.0-2.5$, $3.2- 4.0$ and $6.3-7.9 \times10^{13} \masa$, 
from left to right and top to bottom. 
Vertical dashed lines denote the first and fourth quartile of the distribution.
\label{fig1}}
\end{figure}

Figure \ref{fig1} shows the distribution of formation times for the four subsamples
of haloes as defined previously. The vertical dashed lines denote
the location of the first and fourth quartile of the distribution. 

\subsection{SDSS DR7 Galaxy Groups}
\label{grupossloan}
\begin{figure}
\centering
\resizebox{\hsize}{!}{\includegraphics{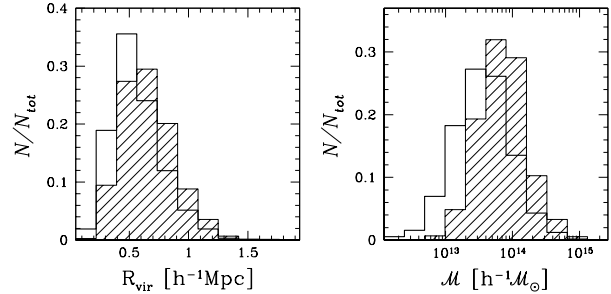}}
\caption{ 
Fractional distribution of virial radii (left panel) and masses (right panel) of
groups in a volume limited sample (with 10 or more members) extracted
from the SDSS DR7 corresponding to  $0.06<z<z_{\rm lim} = 0.10$ (dashed histogram), compared
to that corresponding to the total sample of groups (with 3 or more
members) up to $z=0.10$ (empty histogram). 
\label{fig2}}
\end{figure} 

In recent years a considerable effort has been invested in identifying, using a
multitude of methods, groups and clusters of galaxies in redshift surveys
(e.g. Huchra \& Geller 1982; Tully 1987;
Nolthenius \& White 1987; Ramella et al. 2002; Merch\'an \& Zandivarez
2002, 2005; Gal et al. 2003; Gerke et al. 2004; Lee et al. 2004;
Lopes et al. 2004; Eke et al. 2004; Tago et al. 2006, 2008;
Berlind et al. 2006; Crook et al. 2007; Yang et al. 2008; Sharma \&
Johnston 2009; Finoguenov et al. 2009).

In this work we identify groups of galaxies in the SDSS DR7 
spectroscopic galaxy sample, which comprises of more than 900000 galaxies,
within an area of approximately 9000 square degrees, and with a
limiting r-band magnitude of $m_{\rm lim}=17.77$. 
The group finding algorithm is based on the procedure detailed in
Merch\'an \& Zandivarez (2005) and consists in using
the friends-of-friends algorithm, similar to that developed by
Huchra \& Geller (1982). The algorithm links galaxies ($i$ and $j$) which satisfy
$ D_{ij} \leq D_0 R(z) $ and $ V_{ij} \leq V_0 R(z)$, where $D_{ij}$ is
their projected distance and $V_{ij}$ is their line-of-sight velocity
difference.
The scaling factor $R(z)$ is introduced in order to take into account
the decrement of the galaxy number density due to the apparent magnitude
limit cutoff. We have adopted a transverse linking length $D_0$ corresponding
to an overdensity of $\delta\rho/\rho=80$ and a line-of-sight linking length
of $V_0=200 \kms$. 

In order to avoid significant discreetness effects in the group shape and
dynamics determination, we limit our catalogue to those groups with
more than ten members ($n_m\ge 10$).
To further avoid artificial redshift-dependent effects (eg., Frederic 1995; Plionis
et al. 2004; Plionis et al. 2006) and to build a roughly volume
limited sample, we select those groups with 
$0.06<z<0.1$ and $M_{10}<M_{\rm lim}$, where $M_{10}$ is the absolute magnitude
of the $10^{th}$ brighter galaxy and $M_{\rm lim} (=-19.53)$ is the absolute magnitude 
of a galaxy with apparent magnitude $m=m_{\rm lim}$ at $z=0.1$.
The resulting sample, within the previously mentioned $z$-range,
comprises 760 groups with at least ten member galaxies.

For these groups, we compute their virial radius, 
line of sight velocity dispersion ($\sigma_z$) and virial mass ($\cal M$), following 
Merch\'an \& Zandivarez (2005). 
In Figure \ref{fig2} we present the distribution of virial
masses and radii of the resulting groups, comparing with those of the whole
($n_m\ge 3$ and $z<0.1$) parent group sample. It is evident that we are preferentially selecting 
the richest systems, which is the price we have to pay for producing a
volume limited subsample.

\subsubsection{Group Projected Shape}
The group elongation is determined by diagonalizing the 
''inertia" tensor, which we construct weighting each galaxy $n$ 
by its luminosity $L_n$. This is done in order to weight more the
galaxies that dominate in shaping the group gravitational potential:
\begin{equation}
I_{ij} = \sum_{n=1}^{10} L_n~x_{i,n}~x_{j,n},
\end{equation}
where $x_{i,n}$ is the $i^{th}$ component of the position vector of the
$n^{th}$ galaxy relative to the group center of mass.
Diagonalizing the $I_{ij}$ tensor we obtain two eigenvalues which are related
to the square root of the minor and major axes, $b$ and $a$
respectively, which define the halo axial ratio: $q=b/a$.
\begin{figure}
\centering
\resizebox{\hsize}{!}{\includegraphics{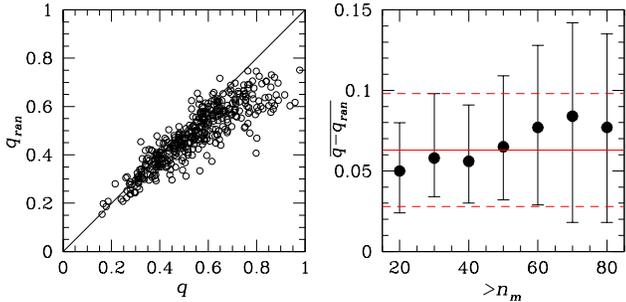}}
\caption{
Left panel: The $q-q_{ran}$ scatter plot for groups within the volume
limited sample and with at least 20 member galaxies.
Right panel: The median value of $q-q_{ran}$ as a function of minimum
number of group galaxy members. Errorbar correspond to the 33\% and
67\% quantiles. The solid line corresponds to the value
$q-q_{ran}=0.063$ and the dashed lines to the $\pm 0.035$ range.
\label{erroresforma}}
\end{figure}

Furthermore, as it is well documented, there exists a dependence of
the group halo apparent elongation on the member-number resolution 
of the system (eg., Paz et al., 2006; Plionis, Basilakos \& Ragone-Figueroa, 2006).
According to this effect, a given system will appear artificially more
elongated (smaller $q=b/a$ ratio)  when sampled by a lower number of
members.
This effect however is large for groups with less than ten members
(see fig.3 in Plionis et al. 2006) and therefore it should not affect
significantly our group sample, which by construction has $n_m\ge
10$. In any case and to overcome any residual effect which could be
introduced due to the variable resolution of the different groups,
we sample each group having $n_m>10$ by selecting randomly only ten galaxies, which
is the number resolution of the lowest $n_m$ galaxy groups in our
sample. Finally, the adopted value
of the projected shape, $q_{ran}$, is the result of computing the mean of 10 realizations of
the random galaxy selection procedure. {\em By this procedure we degrade
the accuracy of the shape parameter of groups with $n_m>10$, but we
gain resolution consistency over all our group sample}.
To get an idea of the uncertainty introduced by our procedure, we
present in the left panel of Figure \ref{erroresforma} the $q$ vs. $q_{ran}$ 
correlation. As expected, the $q_{ran}$ values are systematically lower than their 
counterparts, $q$, computed using all member galaxies, while the deviation increases for
the intrinsically more spherical systems. Since in this plot we show
together all groups with $n_m\ge 20$, in the right panel of
Fig.\ref{erroresforma} we plot the median value of $q-q_{ran}$ as a
function of minimum number of group member galaxies. It can be seen
than there is, as expected, a tendency of $q-q_{ran}$ to slowly increase
with $n_m$ but, within the errors, we can quote an overall median
value of:
$\overline{q-q_{ran}}\simeq 0.063 \pm 0.035$, independent of $n_m$.

\begin{figure*}
\centering
\resizebox{14cm}{12cm}{\includegraphics{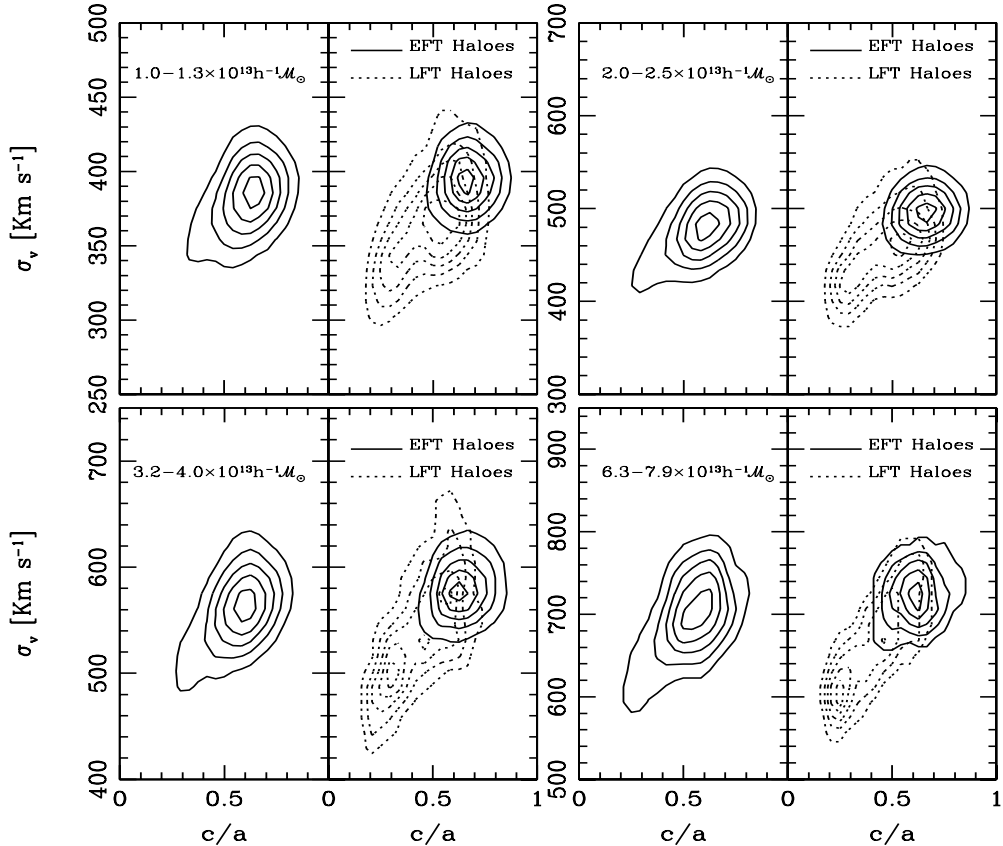}}
\caption{ 
Each panel corresponds to haloes in the four different ranges of mass,
increasing in mass
for the top left, to top right, then bottom left and bottom right panels, respectively. 
The left box of each panel shows the isodensity contours of
$c/a$ vs $\sigma_v$ for all haloes in the corresponding range of mass. 
The right box of each panel shows the corresponding isodensity contours but only of the 
25\% and 75\% quantiles of the formation time distribution, which 
correspond to the EFT (solid line) and LFT (dotted line) families of haloes.
}
\label{shapesig}
\end{figure*} 

\section{THE HALO SHAPE-DYNAMICS RELATION}
\label{hsdr}

In this section we investigate the relation between morphology
and dynamical status of DM haloes.

As it is well known halo properties can have a strong dependence on halo
mass. Especially, the velocity dispersion of virialized or
nearly virialized haloes are related to the halo mass through the virial
theorem: $M\propto \sigma^2$. Therefore, in order to search for an intrinsic dependence of
the velocity dispersion on shape, it is
imperative to suppress their possible dependence to mass, due to the
virialization expectation.
To this end we have 
studied the shape-$\sigma_v$ correlation, in the four mass ranges
defined in section (2.1.2), both in 3D-space and in
projected 2D-space (using 1D velocities and projected halo
shapes).

\subsection{3D Halo $c/a-\sigma_v$ Correlation}
\label{relacion3d}
Figure \ref{shapesig} shows the isodensity contours of the shape-$\sigma_v$ 
scatter plot, at redshift $z=0$, for some of the selected mass bins
(left boxes of each panel).
As it can be seen, there exists a clear correlation between halo
sphericity and halo velocity dispersion, according to which more spherical haloes
have higher velocity dispersions.
Even though we have been cautious in selecting small mass ranges, 
the observed behavior could, in principle, be attributed to the halo 
mass variation within each mass-bin.
For this to be true the more massive systems within each bin
- which should also be the higher velocity dispersion ones - should have 
the highest values of the $c/a$ ratio. However, it is well established that
more massive systems tend to be less spherical (e.g. Allgood et
al. 2006). Therefore, the observed correlation cannot be attributed
to a possible residual dependence on halo-mass. A rather interesting possibility is
related to the fact that haloes, of any
given mass and at any given cosmic epoch, could have a different formation
time as well as a different evolutionary history.

Under this hypothesis, and in order to further investigate the 
dependence of shape-$\sigma_v$ correlation on the halo formation
time ($z_{\rm form}$), we plot separately the EFT and LFT $c/a-\sigma_v$
correlations in the right boxes of each panel of Figure
\ref{shapesig}. It is evident  
that two populations occupy clearly delineated regions in the $c/a-\sigma_v$
plane, with the EFT haloes (solid line) typically having higher velocity dispersions and 
higher sphericities than the LFT haloes (dotted line). 
It can be also observed that
the LFT contours cover a wider range of $\sigma_v$ values than the
corresponding EFT ones, but they do not reach the highest EFT
sphericities.
Finally, it is interesting to also note that there is a shift of the
peak of the LFT contours towards lower $c/a$'s and $\sigma_v$'s 
between the lower mass (upper panels) and the higher mass LFT halos (lower panels). This
could be attributed to a fraction of the most massive LFT systems being relatively more 
dynamically young and in the phase of primary mass aggregation. We will return to this issue 
later on.

As it is obvious from Figure \ref{shapesig}, the same overall 
trend, between the LFT and EFT halos, is present
in all mass bins. It would be ideal then, in order to enhance the
statistical validity of our results, to combine the 
information contained in the different mass ranges in a unique,
independent of halo mass, analysis. This would also facilitate the
corresponding study of the observational data, since the available
number of SDSS groups is significantly smaller than that of the simulated
haloes and thus we cannot afford to divide them in many mass subsamples.

Since the parameters under study ($\sigma_v$, $c/a$ and
$z_{\rm form}$) depend on halo mass,
we have devised a normalization procedure that imposes the mass
independence of the results and combines in a unique relation the
halo velocity dispersion and shape of the whole sample.
In detail we derive normalization functions of the different
parameters by computing
analytic fits of the relation between the corresponding parameter ($P$)
and the halo mass (${\cal M}$). Then for a given halo mass 
we normalize each of the three parameter under study to the value
expected from the analytic fit ($\overline{P}({\cal M})$).

\begin{figure}
\centering
\resizebox{\hsize}{!}{\includegraphics{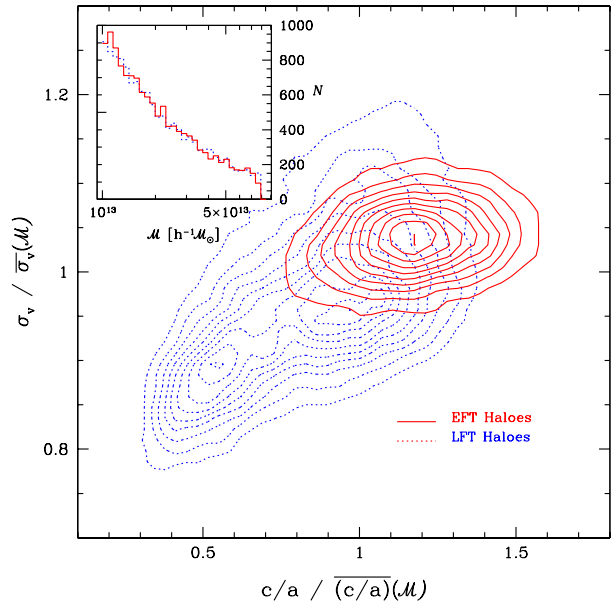}}
\caption{Normalized $c/a$-$\sigma_v$ relation for EFT (solid line) and LFT (dashed
line) populations. All masses are considered. The formation time used to
divide EFT and LFT is also normalized to the expected value for a given
mass (see text).
{\em Inset panel}: Distribution of masses of the EFT (solid line) and LFT (dashed
line) populations demonstrating that no mass
effect is present in the result obtained in the main panel. 
\label{isosgott}}
\end{figure}

Since it is no longer necessary to split haloes in narrow mass bins we can use 
the entire main sample of haloes to plot in Figure \ref{isosgott} the normalized,
independent of mass, shape-dynamics correlation.
It is evident that the EFT and LFT behavior, seen in
Figure \ref{shapesig}, is reproduced for the whole halo sample and we
have further verified 
that these results are indeed independent of halo
mass. In the inset of Figure \ref{isosgott} we show that the
corresponding distribution of masses of the EFT (solid line) and LFT (dashed line)
populations are statistically equivalent and therefore
there is no residual mass-dependent effect affecting the correlation shown in the main panel.

\begin{table}
\caption[]{The Spearman coefficient (SC) for the velocity dispersion-shape correlation 
for haloes in four mass bins. 
In all cases (3D and 2D) the correlations of
the EFT haloes is significantly lower than that of the LFT haloes.}
\vspace{0.03cm}
\tabcolsep 10.5pt
\begin{tabular*}{8cm}{lcccc}
\hline
Sample &\multicolumn{4}{c}{Mass [$\times 10^{13} {\cal M}_{\odot}h^{-1}$]}\\
       & 1.0-1.3& 2.0-2.5& 3.2-4.0&6.3-7.9 \\ \hline \hline
3D All &  0.37  &  0.42  &  0.44  & 0.48  \\ 
3D LFT &  0.45  &  0.55  &  0.60  & 0.72  \\ 
3D EFT &  0.11  &  0.16  &  0.21  & 0.22  \\ \\
2D All &  0.27  &  0.31  &  0.33  & 0.34  \\
2D LFT &  0.31  &  0.38  &  0.42  & 0.47  \\
2D EFT &  0.11  &  0.12  &  0.15  & 0.18  \\
\hline
\end{tabular*}
\label{spearman}
\end{table}

Our main results, as extracted from Figures 4 and 5, can be tabulate 
as follows:
(a) The LFT haloes show a significantly stronger $c/a-\sigma_v$ correlation than the EFT haloes, while they
also span a wider range of values,
(b) The LFT  $\sigma_v$ values are, in some cases, even higher than those of 
the EFT ("virialized") halo population,
(c) A bimodal pattern is present in the $\sigma_v-c/a$ plane (Figure \ref{isosgott})
with two local maxima, one 
of which is situated at the low velocity dispersion and low
$c/a$ values, which as we have discussed earlier and deduced from Figure \ref{shapesig},
it is caused by the more massive haloes. Below we attempt to interpret individually 
each of the above results:

(a) {\em Different $\sigma_v-c/a$ correlation strength of the EFT and LFT haloes:}
The apparent lack of a strong correlation of the EFT halo population
should be attributed to the fact that EFT haloes had plenty of time to
virialize and that the virialization process erases
the intrinsic shape-$\sigma_v$ correlation.
To quantify the strength of the shape-$\sigma_v$ correlation we estimate the  
Spearman correlation coefficient as a function of halo mass (Table
\ref{spearman}). Clearly the observed strong and significant correlation of the
whole halo population should be attributed to the LFT haloes,
since (as we already discussed) the EFT haloes show very weak
shape-$\sigma_v$ correlations.

So far, the results obtained suggest that the intrinsic halo shape-$\sigma_v$ correlation,
within each small mass range, is the consequence of
the coexistence of haloes of different formation times, going through different 
evolutionary phases\footnote{Note that a similar conclusion was reached in a recent
observational study of $z\mincir 0.04$ groups of galaxies (Tovmassian \& Plionis 2009).}.
Dark matter haloes, in the initial stages of formation and
before virialization is complete, have typically a smaller velocity dispersion
and a more elongated shape. Once virial equilibrium is reached
these quantities stabilize and, since this equilibrium configuration is more likely 
to be achieved by the early formation time haloes, the EFT family shows
only a weak such shape-$\sigma_v$ correlation.

\begin{figure}
\centering
\resizebox{\hsize}{!}{\includegraphics{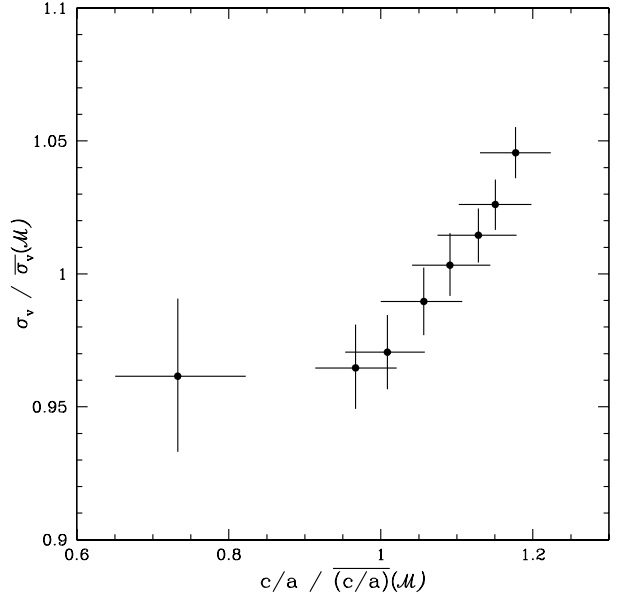}}
\caption{
Normalized $c/a-\sigma_v$ trend corresponding to the entire sample
of haloes. 
Each point corresponds to the median $c/a/(\overline{(c/a)}({\cal M}))$ and 
$\sigma_v/(\overline{\sigma_v}({\cal M}))$ values 
computed for haloes with $z_{\rm form}$ within different octiles of formation time.
Error bars mark the 10\% percentile of the corresponding distribution.}
\label{edadesgott}
\end{figure}

We attempt to investigate in more detail the evolution paradigm, as the
cause of the $\sigma_v-c/a$ correlation, by dividing the halo
formation-time distribution in eight quantilies.
We then compute the median values of $\sigma_v/(\overline{\sigma}({\cal M}))$ and 
$c/a/(\overline{(c/a)}({\cal M}))$ within each
$z_{\rm form}$ quantile and plot them in Figure \ref{edadesgott}.
In concordance with Figure \ref{isosgott}, we consistently find a clear
trend indicating that the earlier the halo is formed, the higher its $\sigma_v$ 
and $c/a$ ratio, as expected from the fact that older haloes have more time to evolve and virialize.
Evidently, a halo evolutionary sequence appears, 
with dynamically young haloes situated at the lower-left of the 
$c/a-\sigma_v$ plane  
and as they evolve, they move towards higher $\sigma_v$ and $c/a$ values, 
where the virialized haloes are located.

Note, however, that in Figure \ref{edadesgott} the dynamically youngest haloes 
(1st octile of formation time)
show a behaviour that deviates from the overall trend, being characterized by a 
lower normalized $c/a$ value.  
This outlier has a relatively larger uncertainty with 
respect to the other octiles, and a possible cause of this behaviour is presented in (c).

(b) {\em LFT haloes with $\sigma_v$ values even larger than those of EFT haloes:} 
In an attempt to explain what is the cause of the rather unexpected behavior
of some LFT haloes, i.e., those having relatively high sphericity and
in the same time
a velocity dispersion which is as large or even larger than what expected from the
``virialized'' EFT haloes, we suggest two mechanisms:
\begin{itemize}
\item Since LFT haloes are prone to have had a recent merger and 
  since it is expected that dynamical interactions and 
  merging processes increase the velocity dispersion of the involved systems
  some merging or highly interacting haloes could be undergoing a
  transient geometrical configuration of high sphericity (resulting
  when the in-falling substructures are near the perigee), while
  temporarily achieving a velocity dispersion which could be even larger than their 
  virial value (Faltenbacher et al. 2006).
  We have verified that indeed this is the case for some
  of the LFT haloes. 
\item  Part of the high velocity dispersion and quasi-spherical LFT
  haloes could be due to truly virialized LFT haloes by
  a somewhat rapid virialization process depending on their particular
  formation history and/or environment. 
\end{itemize}

\begin{figure}
\centering
\resizebox{\hsize}{!}{\includegraphics{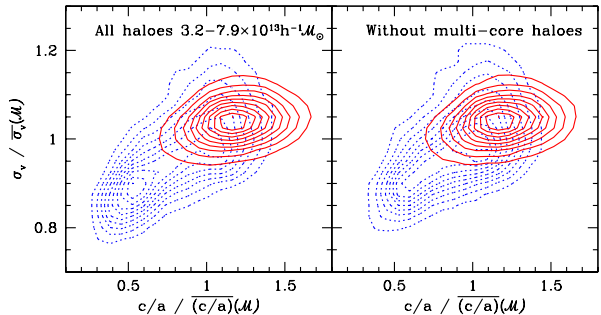}}
\caption{ Same as Figure \ref{isosgott} but only considering haloes
with masses between  $3.2$ and $7.9 \times 10^{13} \masa$. In the left panel the entire sample
is used whereas in the right panel all double-core haloes have been excluded.}
\label{double-core}
\end{figure} 

(c) {\em The two maxima in the LFT $\sigma_v-c/a$ correlation plane:}
As we have discussed earlier, the lower $\sigma_v-c/a$ maximum seen in 
Figure 5 is caused by the most massive haloes (ie., 
those with ${\cal M} \magcir 3.2 \times 10^{13} \;h^{-1} \; {\cal M}_{\odot}$) and 
could be attributed to a fraction of the most massive LFT systems being relatively 
dynamically younger and in the phase of merging, where the FOF algorithm could join a merging pair 
into a single FOF halo.

In order to verify our suspision
we have selected haloes with masses
between $3.2 - 7.9 \masa$ (including the two most massive ranges presented
in Figure \ref{shapesig}, where the bimodal pattern was evident for the LFT haloes) and 
investigated in which manner the described trend would be 
affected if double (or multiple)-core halos (or otherwise dumbbell shaped haloes) were excluded.
Double-core haloes were identified as those haloes that are split in two or 
more subhaloes of at least 1000 particles, within a relative distance of 2$r_{\rm sph}$ 
($r_{\rm sph}$ is the diameter of a sphere which has the same volume as the
considered halo),
when a re-identification with a shorter linking length 
(0.135 of the mean interparicle separation) is performed. Therefore, they are 
major pre-merger systems which the FOF joins into a single system and which
other halo-finders would have probably singled out each component as an individual halo.
In the left panel of Figure \ref{double-core} we show the $c/a-\sigma_v$ correlation
when all haloes in the mentioned mass range are considered, whereas in the right panel
double-core haloes are excluded.
It is evident that the secondary maximum at low $\sigma_v$ and $c/a$, 
is caused by the
double-core haloes, which are those that are in an active major merger process, 
and thus an integral part of the ``dynamically young'' family of haloes.
Note, that although the $c/a-\sigma_v$ LFT correlation is weakend when
excluding the double (multiple)-core haloes, it is still clearly present.

Returning to the issue of the ``outlier'' seen in Figure \ref{edadesgott}, 
we have verified that the cause
of the significantly smaller normalized $\langle c/a \rangle$ value of the first 
formation time octile is the significantly larger number of double 
(multiple)-core haloes found in this octile with respect to the other 
octiles. For example, for the case of the most massive haloes (ie., those investigated 
in Figure \ref{double-core}) we find that $\sim 22\%$ of the haloes in the first 
octile are double (multiple)-core haloes, while this number drops to $4.8\%$ 
and $2.6\%$ in the second and third octile respectively, and to even smaller 
values thereafter.

\begin{figure}
\centering
\resizebox{\hsize}{!}{\includegraphics{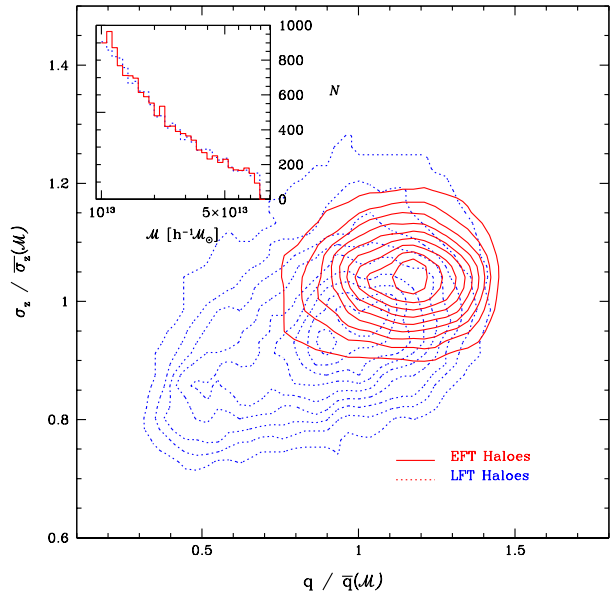}}
\caption{
As in Fig.\ref{isosgott}  but using projected halo shapes and 1D velocity dispersions.}
\label{proy}
\end{figure}

\subsection{2D Halo $q-\sigma_{z}$ Correlation}
\label{proysection}
Here we present the corresponding analysis of the previous
subsection, but using the projected halo axial-ratio ($q=b/a$ obtained
from the projected particle distribution) and the
one-dimensional velocity dispersion. This is done in order 
to see how the previous (3D) results are translated into theoretical
expectations that can be compared with the results based on 
observational group samples.

Figure \ref{proy} is the 2D analogue of Figure \ref{isosgott} and we can clearly see
that qualitatively the correlation revealed in Figure \ref{isosgott}, albeit weaker, 
is repeated in Figure \ref{proy}. Details and significances may differ 
(see Table \ref{spearman} for different mass ranges), 
however, the main result is that we expect to see (in realistic group samples) 
hints of an independent of mass correlation of the projected halo axial-ratio with its
line-of-sight velocity dispersion.

\section{A Formation Time Proxy}
\label{proxies}

The aim of this section is to provide a formation-age proxy,
tested with simulated haloes, but that it can also be
computed in realistic groups of galaxy samples, where the adopted
definition of $z_{\rm form}$ is not usable. 
This is done in order to
test the morphology-dynamics-formation time correlation
(which was found  to be present in simulated haloes)
in groups extracted from the SDSS DR7.

We propose as a halo formation-age proxy ($\Delta_{\rm subh}$), 
the fraction of mass that is
associated to the most massive substructure (subhalo), when a further
re-identification is performed with a linking length which equals 
half the original one ($0.17/2$ of the mean inter-particle separation). 
This proposed parameter can be interpreted as a
concentration indicator since it accounts for the distribution of
mass at two different overdensities (given by the used linking lengths). 

\begin{figure}
\centering
\resizebox{\hsize}{!}{\includegraphics{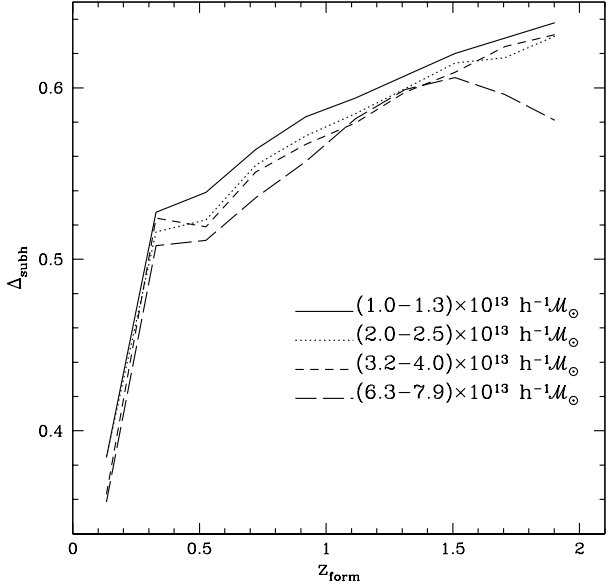}}
\caption{
The $\Delta_{\rm subh}$ parameter as a function of formation time
computed for dark matter haloes in four ranges of halo mass.
\label{sub-zform}}
\end{figure}

The number of particles of a halo in each overdensity level 
is defined as $n_{\rm high}$ and $n_{\rm low}$ for the higher and lower 
overdensities, respectively.
Then the concentration parameter is given by:
\begin{equation}
\Delta_{\rm subh} =  \frac{n_{\rm high}}{n_{\rm low}} \;.
\end{equation}
The parameter $\Delta_{\rm subh}$
takes values within the range $\Delta_{\rm subh} \in [0, 1]$, with a
value near 1 indicating a high concentration and 
a value near 0 indicating the opposite. 
We show in Figure \ref{sub-zform} the relation between $\Delta_{\rm subh}$ and 
halo formation redshift for the different ranges of mass labeled in the plot. 
It is evident that the oldest haloes have the highest levels of 
concentration, whereas
the youngest ones extend to lower values of $\Delta_{\rm subh}$.

Since the above analysis clearly demonstrates that the proposed parameter
can be used as an age indicator, we continue  
with the computation of the concentration parameter corresponding to galaxy groups in the 
observational data, $\Delta_{\rm sub}$.
To this end we
perform a second friends-of-friends identification, but this time
using an overdensity of $\delta \rho/\rho=300$. This procedure allows us to extract 
a denser subgroup, of a given system, with respect to the original
$\delta \rho/\rho=80$ sample. 
Then $\Delta_{\rm sub}$ is given by the ratio between the number 
of members of the denser subgroup to the number of members of its
parent ($\delta \rho/\rho=80$) group. Utilizing  this formation-age proxy,
we examine the SDSS-DR7 group shape-dynamics-formation time relation 
in the next section.

\section{The Shape-Velocity dispersion Relation of SDSS DR7 Groups}
\label{2d}
Many studies have attempted to determine the morphology and
dynamical status of groups of galaxies (eg. Hickson et al. 1984; Malykh \& Orlov
1986; Orlov, Petrova, Tarantaev 2001; Kelm \& Focardi 2004, 
Plionis, Basilakos \& Tovmassian 2004; Plionis, Basilakos, \&
Ragone-Figueroa 2006;  Da Rocha, Ziegler, \& Mendes de Oliveira 2007; 
Coziol \& Plauchi-Frayn 2007; Wang et al. 2008; Hou et al. 2009) and have found that 
their morphology corresponds to that of mostly prolate-like triaxial ellipsoids
(see however Robotham, Phillips \& de Propris 2007), while there also
appears to be a correlation between velocity dispersion and projected
group shape (eg. Tovmassian \& Plionis 2009) similar to what we found
in the previous section. 

Here we analyse our ``volume-limited'' group catalogue based on the SDSS DR7
galaxy survey, which was described in section \ref{datos}. 

\begin{figure}
\centering
\resizebox{\hsize}{!}{\includegraphics{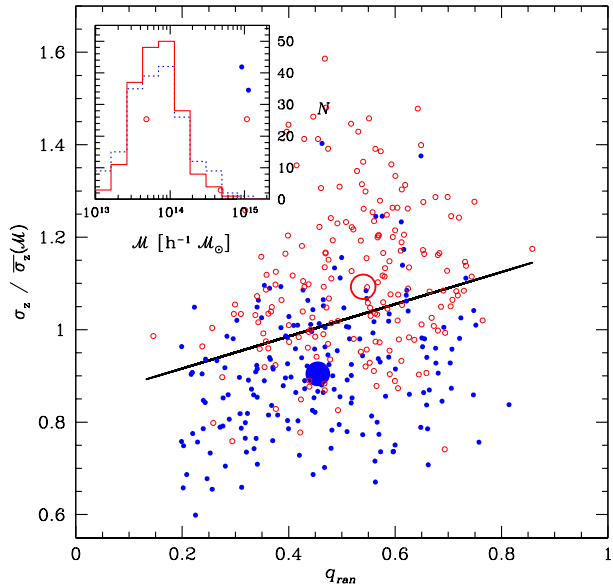}}
\caption{ 
$q_{ran}$ vs. normalized $\sigma_z$ relation for groups in the SDSS DR7.
Filled (low concentration) and empty (high concentration) dots stand for 
groups in the first and last quartile 
of the normalized $\Delta_{\rm sub}$ distribution, respectively. The large circles
correspond to the median values. 
Solid line shows the least square
fit using all the groups, irrespective of their $\Delta_{\rm sub}$ value.
The inside box shows the mass distributions for groups in the cuartiles of
low and high concentration (dotted and solid lines respectivelly).
\label{puntosgott}}
\end{figure}

As we have concluded in section \ref{proysection}, the shape-$\sigma$ correlation 
for dark matter haloes, extracted from the simulation, survives in
projection but it is significantly weakened when the projected 
shape ($q$) and  the one dimensional velocity dispersion are used. 
Further scatter should be expected in a more realistic situation in which 
the projected shape of SDSS groups are computed using only ten members
($q_{ran}$). 
Indeed, as it was shown in section \ref{grupossloan},  
the typical error, expected when only ten members are considered, 
in the estimation of the projected shape is non-negligible. 
Such issues may act as to hide the morphology-dynamics correlation of 
observed groups of galaxies, and thus even a weak such correlation
should be considered as a hint of a true underlying effect.  
We show in Figure \ref{puntosgott} the least square best fit line of the 
$q_{ran}$ vs. $\sigma_z/\bar{\sigma_z}({\cal M})$ scatter plot 
(of the entire sample of galaxy groups)
where the observed trend is in full agreement with
Figure \ref{proy}. 

We now attempt to investigate whether the observed 
$q_{ran}$ vs. $\sigma_z$ correlation,
in Figure \ref{puntosgott}, is due to the coexistence of groups of different formation
times, as was indeed shown to be the case in the dark matter halo analysis.
As shown in section 3.2 (see also Figure \ref{proy}), using projected shapes and
one dimensional velocity dispersions does not completely mask the
systematic behaviour observed for haloes of different
formation times. However, taking into account 
the unfeasibility of determining the formation time of the observed galaxy groups, 
we use as a proxy of the group dynamical age its concentration
parameter, $\Delta_{\rm sub}$. This parameter has indeed been shown (for the case of DM
haloes) to correlate with halo formation time (see section \ref{proxies}).

In Figure \ref{puntosgott}  we separately plot groups of the first and last quartile 
of the $\Delta_{\rm sub}$ parameter distribution (as was done for formation time in 
section \ref{relacion3d}).
Filled and empty circles represent galaxy groups 
with low and high concentration (corresponding to groups in the first and
last quartiles of the $\Delta_{\rm sub}$ distribution), respectively. 
The tendency for high concentrated groups (open circles) to have systematically 
higher velocity dispersions and $q$ values is quite evident. We also plot their
corresponding median values as large filled and empty circles, from
which it can be seen that indeed there is a separation of
the two families (based on $\Delta_{\rm sub}$), in both $q_{ran}$ and $\sigma_z$ axes.
We have again verified that the observed behaviour is not due to 
a mass-dependent effect, since the
distribution of mass of the two group families are statistically
equivalent (see insert of Figure  \ref{puntosgott}).

It is important to keep in mind that the uncertainty in 
the determination of the projected group shape, 
induced by our random sampling procedure aiming in
reducing the variable resolution effects, 
could introduce a large scatter in the morphology-dynamics
correlation. This could then act in the direction of hiding the
differences between the two families of groups, represented by the
different symbols in Fig.\ref{puntosgott}. 
We have tested this assertion by using, within small ranges of
group mass and for a given number of galaxy members, 
all the available galaxy members to estimate more accurately the group $q$ values, 
in the expense however of working with a low number of groups. 
We now find significantly larger differences between 
the mean $q$-values of the low and high concentration families. 
Therefore, had we higher resolution measurements of group shapes,
we would have recovered a significantly larger separation between the low and high
concentrated group families in the $q-\sigma_z$ plane,
strengthening the interpretation of group formation time differences being the
cause of the observed $q-\sigma_z$ correlation, as is the case in the
dark matter haloes analysed.

\section{SUMMARY \& CONCLUSIONS}
\label{conc}

We have identified dark matter haloes in the {\sc MareNostrum} Universe
and groups of galaxies in the SDSS DR7 with the purpose of
searching for possible, independnet of mass, 
dependencies of halo/galaxy groups shape on their dynamics.

We have computed the halo formation time as the redshift at which the 
halo accretes half of its final mass. Since this definition of formation 
time is not applicable to observed groups identified in realistic galaxy
catalogues, we have identified a concentration parameter that can be
used as a proxy of the formation time of haloes and at the same time 
it is easily computed for the observed SDSS galaxy systems.

On this regard we have found a significant correlation between halo
shape and dynamics 
(velocity dispersion) which is independent of the halo mass. We have
identified the cause of this correlation to be the halo formation time 
differences. 
Early formation-time haloes show a very weak such correlation, 
being more spherical, having a higher velocity dispersion and
showing substantial concentration. On the contrary, late 
formation-time haloes show a strong shape-dynamics correlation (having
a wide range of shape and velocity dispersion values),
with typically lower velocity dispersions, lower sphericities and 
significant lower concentrations.
We have also studied the influence of multi-core haloes, i.e.,
major merger systems which the FOF joins into a single system and which
other halo-finders would have probably singled out each component,
in the strength of the LFT shape-dynamics correlation and verified that 
although they enhance the correlation, it however persists even excluding
this subsample of haloes.

Finally, we find a fraction of the LFT haloes showing a somewhat 
unexpected behaviour, having high sphericities and velocity dispersions 
(comparable or even higher than those of more virialized systems), a fact which could
be attributed to either
some late formation-time haloes having transitory spherical shapes 
and large velocity dispersion, during the minimum pericentric passage of 
a merging process, or/and to a relatively faster virialization process
occuring in some of these haloes.

The halo morphology-velocity dispersion correlation, albeit weaker, 
survives the 2D projection, indicating that an analogous behaviour should
be expected in observational group samples.
Applying a similar analysis to a ``volume-limited'' 
subsample of the SDSS DR7 groups of galaxies,
identified using a friends of friends algorithm, we have also found a
group shape and dynamics correlation, which is independent of group mass (see also 
Tovmassian \& Plionis 2009).
Using a concentration parameter as a proxy of the formation time, we indeed  
find a significant, independent of mass, shape-dynamics correlation, in a similar fashion 
as in the dark matter halo case. Less concentrated (late forming) groups 
are typically less spherical and have lower velocity dispersion than their 
equal mass more concentrated (early forming) counterparts.

\section*{Acknowledgments}
This work has been partially supported by the European Commission's 
ALFA-II programme through its funding of the Latin-american European Network 
for Astrophysics and Cosmology (LENAC), 
the Consejo de Investigaciones Cient\'{\i}ficas y T\'ecnicas de la Rep\'ublica 
Argentina (CONICET) and the Secretar\'{\i}a de Ciencia y T\'ecnica de la 
Universidad Nacional de C\'ordoba (SeCyT).

The {\sc MareNostrum Universe} simulation has been performed at the Barcelona
Supercomputer Center  and  analyzed at NIC J\"ulich supercomputer center.
GY would like to thank also MCyT for financial support under
project numbers  FPA2006-01105  and AYA2006-15492-C03.

\end{document}